\documentclass[conference]{IEEEtran}

\input{glyphtounicode.tex} 
\pdfmapline{+ptmr8r Times-Roman "TeXBase1LetterEncoding ReEncodeFont" <8r.enc <ptmr8a.pfb} 
\pdftrailerid{}

\IEEEoverridecommandlockouts
\usepackage{cite}
\usepackage{amsmath,amssymb,amsfonts}
\usepackage{algorithmic}
\usepackage{graphicx}
\usepackage{xcolor}
\def\BibTeX{{\rm B\kern-.05em{\sc i\kern-.025em b}\kern-.08em
    T\kern-.1667em\lower.7ex\hbox{E}\kern-.125emX}}

\usepackage{listings}
\usepackage{xcolor}
\usepackage{multirow}
\usepackage{enumitem}
\usepackage{url}
\usepackage{booktabs}
\usepackage{newtxtext,newtxmath}
\usepackage{pifont}

\newcommand{\cmark}{\ding{51}} % ✓
\newcommand{\xmark}{\ding{55}} % ✗

\definecolor{frameworkcolor}{HTML}{52175b}

\lstdefinestyle{cppcode}{
  language=C++,
  basicstyle=\ttfamily\footnotesize,
  keywordstyle=\color{blue},                            
  morekeywords=[4]{void, override, final, int64_t, uint64_t, nullptr, static_assert, constexpr, noexcept, INT64_MAX, InitializeStateToInt64Max},
  morekeywords=[3]{vid_t, GetValue, SetValue, GetInnerOutEdges, GetOuterInEdges, HasVertex, SetVertex, InitializeStatesTo, GetUpdatedVertices, WorkerCtx, DEFINE_WORKER_CLASS, END_WORKER_CLASS, DEFINE_ALGORITHM, END_DEFINE, REGISTER_ALGORITHM, GetEdges, GetVertexId, VertexRange, InitializeStateAsId},
  keywordstyle=[3]\color{frameworkcolor},                       
  numbers=none,
  stepnumber=1,
  breaklines=true,
  frame=single,
  showstringspaces=false,
  columns=flexible,
}
\begin{document}

\title{GraphFlash: Enabling Fast and Elastic Graph Processing on Serverless Infrastructure}

% \author{\IEEEauthorblockN{Chen Zhao}
% \IEEEauthorblockA{\textit{DisNet Lab.} \\
% \textit{Computing and Information Systems}\\
% \textit{The University of Melbourne}\\
% Melbourne, Australia \\
% ch3nzhao@gmail.com}
% \and
% \IEEEauthorblockN{Parsa Poorsistani}
% \IEEEauthorblockA{\textit{DisNet Lab.} \\
% \textit{Computing and Information Systems}\\
% \textit{The University of Melbourne}\\
% Melbourne, Australia \\
% poorsistani13@gmail.com}
% \and
% \IEEEauthorblockN{Mohammad Goudarzi}
% \IEEEauthorblockA{\textit{Faculty of Information Technology} \\
% \textit{Monash University}\\
% Melbourne, Australia \\
% Mohammad.Goudarzi@monash.edu}
% \and
% \IEEEauthorblockN{Tawfiq Islam}
% \IEEEauthorblockA{\textit{DisNet Lab.} \\
% \textit{Computing and Information Systems}\\
% \textit{University of Melbourne}\\
% Melbourne, Australia \\
% tawfiqul.islam@unimelb.edu.au}
% \and
% \IEEEauthorblockN{Adel N. Toosi}
% \IEEEauthorblockA{\textit{DisNet Lab.} \\
% \textit{Computing and Information Systems}\\
% \textit{University of Melbourne}\\
% Melbourne, Australia \\
% adel.toosi@unimelb.edu.au}
%}

\author{
\IEEEauthorblockN{
Chen Zhao\IEEEauthorrefmark{1}, 
Parsa Poorsistani\IEEEauthorrefmark{1}, 
Mohammad Goudarzi\IEEEauthorrefmark{2}, 
Tawfiq Islam\IEEEauthorrefmark{1}, 
Adel N. Toosi\IEEEauthorrefmark{1}\IEEEauthorrefmark{3}\IEEEauthorrefmark{4}
}
\IEEEauthorblockA{
\IEEEauthorrefmark{1}DisNet Lab., School of Computing and Information Systems, The University of Melbourne, Australia\\
\{tawfiqul.islam, adel.toosi\}@unimelb.edu.au\\
\IEEEauthorrefmark{2}Faculty of Information Technology, Monash University, Australia\\
Mohammad.Goudarzi@monash.edu\\
\thanks{\IEEEauthorrefmark{3}Corresponding author.}
\thanks{\IEEEauthorrefmark{4}Adel N. Toosi is supported by the Australian Research Council (ARC) through funded projects DP230100081 and LP210200213.
}
}
}

\maketitle

\begin{abstract}

Graph processing systems are essential for analyzing large-scale data with complex relationships, yet most existing frameworks rely on statically provisioned clusters, resulting in poor elasticity and inefficient resource utilization under dynamic workloads. Serverless computing offers automatic scaling and fine-grained billing, but existing serverless graph systems suffer from performance limitations due to inefficient state management and high communication overhead through external storage. We present GraphFlash, a fast and elastic graph processing framework built on serverless infrastructure. GraphFlash adopts a subgraph-centric programming model and leverages shared external storage for coordination and communication, enabling stateless, fine-grained function execution. It supports two execution modes: \textit{rotating} mode for resource-constrained environments and \textit{pinned} mode for higher performance when resources are sufficient. To address serverless limitations, GraphFlash introduces system-level optimizations, including partition-aware key aggregation, intra-function partition co-location, and superstep-aware activation. Across multiple graph algorithms and datasets, GraphFlash outperforms existing serverless-compatible systems by up to 127$\times$ in execution time and reduces resource consumption by up to 98\% under higher-resource configurations, while matching the performance of traditional distributed frameworks on large workloads. Even with limited resources, it achieves up to 48$\times$ speedup and 99.97\% cost reduction over prior serverless solutions, demonstrating that GraphFlash makes serverless graph processing practical and performant.

\end{abstract}


\section{Introduction}
Graphs are useful for modeling complex relationships in a wide range of domains, including social networks, the World Wide Web, recommendation systems, and knowledge graphs. To analyze increasingly large and complex graphs, a variety of graph processing systems have been developed, which broadly fall into three categories~\cite{coimbra2021analysis}: \emph{single-machine systems}~\cite{10.1145/2517327.2442530,10.1145/2723372.2735369,10.14778/2809974.2809983}, \emph{distributed systems}~\cite{averyGiraph,carbone2015apache,10.5555/1863103.1863113,fan2021graphscope,10.5555/3026877.3026901}, and \emph{HPC systems}~\cite{gregor2005parallel,10.1177/1094342011403516,pearce2013scaling}.

Among these, distributed systems have become a prominent approach to large-scale graph analytics due to their ability to parallelize computation across multiple commodity machines. Frameworks such as Giraph~\cite{averyGiraph} and GraphScope~\cite{fan2021graphscope} achieve high performance by distributing graph partitions and coordinating iterative computation through supersteps. However, these systems rely on fixed clusters, extensive configuration, and persistent resource provisioning—even when workloads are bursty, irregular, or short-lived. This lack of elasticity leads to resource underutilization and high operational costs and overhead, especially when scaling to a variety of graph sizes and algorithmic demands~\cite{toader2019graphless,10.1145/3620665.3640361}.

Serverless computing presents a promising alternative. By abstracting infrastructure management and enabling automatic scaling and pay-per-use billing, it offers an attractive model for dynamic and irregular workloads. It provides flexibility beyond IaaS: graphs of varying sizes can be processed with appropriately scaled functions under true pay-as-you-execute billing. Serverless has shown success in areas such as data analytics and machine learning~\cite{298681,10.1145/3357223.3362711}, its application to graph processing remains limited due to fundamental challenges in managing state, communication, and execution flow across ephemeral functions. Graph workloads are inherently stateful and iterative, yet functions are stateless, short-lived, and rely on ephemeral storage; as a result, vertex and edge states must be maintained outside the functions, creating heavy reliance on external storage systems. Fine-grained message passing across partitions increases communication overhead, while cold starts and limited intra-function memory hinder scalable performance. As a result, existing systems either sacrifice serverless principles (e.g., FaaSGraph~\cite{10.1145/3620665.3640361}) or face efficiency limitations~(e.g., Graphless~\cite{toader2019graphless}).

To address these gaps, we present GraphFlash, an elastic graph processing framework that runs entirely on serverless infrastructure while achieving performance comparable to traditional distributed systems. GraphFlash adopts a subgraph-centric execution model and introduces two modes, \emph{rotating} and \emph{pinned}, to balance elasticity and performance under varying resource conditions.
Our main contributions are as follows:

\begin{itemize}
  \item \textbf{A fully serverless and elastic graph processing framework:} We design GraphFlash, a practical system that supports stateless, subgraph-centric graph processing with elastic resource usage. It introduces rotating and pinned execution modes to adapt to both resource-constrained and high-performance scenarios.

  \item \textbf{Identification and resolution of key serverless bottlenecks:} We analyze the fundamental limitations of serverless platforms for graph workloads and address these challenges through targeted system-level design mechanisms, most notably enabling intra-function multi-core parallelism, complemented by techniques such as partition-aware key aggregation and superstep-aware activation.

  \item \textbf{Performance analysis across workloads and system configurations:} We evaluate GraphFlash using diverse graph algorithms and datasets on real testbed environments and systematically analyze how factors such as partitioning strategy, execution mode, and concurrency level affect performance and cost-efficiency.
\end{itemize}

%The remainder of this paper is organized as follows. Section~\ref{sec:background} introduces the background and motivation. Section~\ref{sec:programming} presents the programming interface. Section~\ref{sec:design} details the system architecture and execution flow, and Section~\ref{sec:opt} describes key optimizations. Section~\ref{sec:evaluation} provides evaluation results, Section~\ref{sec:related} reviews related work, and Section~\ref{sec:conclusion} concludes the paper.

\section{Background and Motivation}
\label{sec:background}
\subsection{Preliminaries}

\subsubsection*{Distributed Graph Processing}
Distributed graph processing systems divide large graphs into partitions that can be processed across multiple machines or compute units. These systems typically use a synchronous computation model based on supersteps. A superstep represents one complete round of computation where: 1) each compute unit processes its assigned partition; 2) necessary messages are exchanged between partitions; 3) a global synchronization ensures all computations and communications are complete before the next superstep begins.

This iterative process continues until the algorithm converges or a termination condition is met. The superstep model ensures consistency and makes distributed algorithms easier to reason about, though it can introduce synchronization overhead \cite{10.1145/1807167.1807184}.

\subsubsection*{Programming Models}

The programming model defines how developers express graph algorithms for distributed execution. Three main paradigms have emerged:

\noindent 1) \textit{Vertex-centric model:} First introduced by Google's Pregel system \cite{10.1145/1807167.1807184}, this ``think-like-a-vertex" (TLAV) approach processes graphs from the perspective of individual vertices. Each vertex can maintain its own state, process messages from adjacent vertices, update its state based on received messages, send messages to its adjacent vertices.
This model has been widely adopted by systems such as Giraph~\cite{10.14778/2824032.2824077} and GraphX~\cite{10.1145/2484425.2484427} due to its simplicity and intuitive nature.

\noindent 2) \textit{Edge-centric model:} This paradigm, implemented in systems such as X-Stream~\cite{10.1145/2517349.2522740}, Chaos~\cite{10.1145/2815400.2815408}, and PK-Graph~\cite{10.1007/978-3-031-17834-4_9}, focuses on edge-based computations. It is more efficient for certain algorithms and graph structures, especially when edge properties or updates are central, as it enables direct access and modification without vertex-based traversal overhead.

\noindent 3) \textit{Subgraph-centric model:} This model processes subgraphs~(also referred to as partitions) rather than individual vertices or edges. A prominent example of a subgraph-centric model is GRAPE \cite{grape}. This model offers several advantages: 1) reduces communication overhead by processing connected components together; 2) enables more efficient local computations within subgraphs; 3) often achieves higher performance by minimising cross-partition messages. This approach has gained prominence in modern systems (e.g., Gemini~\cite{10.5555/3026877.3026901}, GraphScope~\cite{fan2021graphscope}) due to its superior performance characteristics. Our proposed GraphFlash adopts this model to enhance efficiency.

\subsection{Enabling Graph Processing via Serverless Computing}

Serverless computing offers a promising paradigm shift for scalable and elastic data processing. It eliminates the burden of infrastructure provisioning, provides automatic scaling based on workload demand, and charges users only for the actual resources consumed. These characteristics make it an attractive execution model for graph processing workloads, which are often dynamic, bursty, and irregular in resource usage. In particular, graph algorithms, especially those used in real-world analytics pipelines, exhibit varying compute intensity and memory requirements depending on the graph structure and algorithm phase. Serverless platforms can scale up during intensive phases and scale down automatically when the load drops, thereby improving cost-efficiency and elasticity. Recent successes of serverless frameworks in domains such as machine learning~\cite{298681} and relational query processing~\cite{10.1145/3318464.3380609} further motivate their applicability to graph workloads.

However, realizing serverless graph processing is non-trivial. Graph algorithms are inherently stateful and iterative, requiring external storage to maintain vertex and edge state across stateless function invocations. Fine-grained message passing across partitions introduces significant communication overhead, while cold starts and execution constraints further exacerbate these challenges, making na\"{\i}ve
 adaptations inefficient or impractical. These limitations are evident in prior work: Graphless~\cite{toader2019graphless} maps vertices to stateless functions and relies on external storage, but suffers from high communication overhead and limited scalability. FaaSGraph~\cite{10.1145/3620665.3640361} reduces message latency using shared memory and proxies; however, its dependence on co-located containers and direct memory access violates serverless principles, resulting in fixed resource allocation and a less convincing cost model. In contrast, non-serverless systems such as Giraph~\cite{10.14778/2824032.2824077} and GraphScope~\cite{fan2021graphscope} partially mitigate resource wastage by leveraging shared-resource frameworks like Hadoop and Kubernetes.

This motivates a new approach that embraces the benefits of serverless computing while addressing its unique challenges for graph workloads. In this work, we present GraphFlash, a fully serverless, elastic graph processing framework that delivers both performance and cost efficiency through principled system design and targeted optimizations.

\section{Programming Interface}
\label{sec:programming}

This section presents GraphFlash's programming model and introduces core APIs. These design choices aim to optimize performance, reduce memory usage, and provide flexibility for various graph processing tasks.

\noindent\textbf{Programming Model:}
As stated earlier, GraphFlash follows a subgraph-centric model that iterates through supersteps, processing only active subgraphs or vertices per step. By default, execution terminates when no active subgraphs remain, although users can invoke custom APIs to terminate earlier if needed. Our implementation leverages GRAPE~\cite{grape}, a subgraph-centric engine, adopted by other state-of-the-art frameworks such as GraphScope~\cite{fan2021graphscope}. In practice, each partition contains inner vertices that are updated during computation by each worker, and outer vertices that belong to adjacent partitions and serve as read-only inputs.
%Each worker updates only its inner vertices and treats outer vertices as read-only.
To propagate changes, workers send updates to others whose partitions contain the affected outer vertices.

\noindent\textbf{APIs:}
GRAPE defines two core APIs: (1) \textit{PEval}, an initialization function that computes partial results within each partition of the graph; and (2) \textit{IncVal}, an incremental function invoked in subsequent supersteps to process incoming messages and update the local state. Here, messages refer to the data exchanged between vertices (or nodes) during graph processing. These messages convey information, such as updated vertex values or intermediate results, that are necessary for subsequent computations.

GraphFlash provides two versions for Knative and AWS Lambda.
As an example, Fig.~\ref{fig:wccworker} shows the implementation of AWS Lambda of a WCC worker (Weakly Connected Components). The algorithm is registered as `WCC`, with int64\_t chosen as the value type. Users define PEval and IncVal to initialize states and drive iterative updates, while propagation is implemented in a helper function. The framework also offers initialization routines such as InitializeStateAsId, as well as semantic utilities including GetValue, SetValue, and GetEdges to simplify algorithm development.

GraphFlash uses C++ templates to support different graph types and value types. Whether the graph is directed or weighted is determined by template parameters, enabling efficient specialization without redundant data copies or overhead. For example, one may use integer values for BFS and floating-point values for PageRank.

To add a new algorithm, users only need to provide a C++ source file containing the worker logic. During the build process, this file is compiled into a shared object (.so) library, which is dynamically loaded by the AWS Lambda worker at runtime.

\begin{figure}[ht]
  \centering
  \begin{minipage}{0.95\linewidth}
    \begin{lstlisting}[style=cppcode, basicstyle=\ttfamily\scriptsize]
#include "worker/plugin_support.h"

DEFINE_ALGORITHM(WCC) {}

void PEval() override {
  InitializeStateAsId();
  Propagate();
}

void IncVal() override { Propagate(); }

void Propagate() {
  for (vid_t v : VertexRange()) {
    auto l = GetValue(v);
    for (auto &e : GetEdges(v)) 
        l = std::min(l, GetValue(e.GetVertexId()));
    if (l < GetValue(v)) SetValue(v, l);
  }
}
REGISTER_ALGORITHM("WCC", int64_t, WCC)

    \end{lstlisting}
  \end{minipage}
  \caption{A WCC implementation in GraphFlash.}
  \label{fig:wccworker}
\vspace{-0.3cm}
\end{figure}

\section{System Design}
\label{sec:design}

This section provides a detailed description of the components and execution flow of GraphFlash.

\subsection{Components}

GraphFlash adopts a minimalist design to reduce runtime overhead and streamline execution. To support execution in a serverless environment, all graph processing is performed within stateless functions. Because these functions are ephemeral and cannot communicate directly during execution, an external storage system is required for both data persistence and inter-function communication. GraphFlash defines two types of functions: \textit{coordinators} and \textit{workers}. The external storage system is referred to as \textit{Memory-as-a-Service }(MaaS), an adopted terminology from Graphless~\cite{toader2019graphless}.

\subsubsection*{Coordinator}

The coordinator is responsible for initiating new supersteps and distributing workloads to workers, with each task requiring one coordinator. It continuously polls MaaS to determine whether all workers have completed their tasks and to trigger the next superstep. This design aligns with the principles of serverless computing, allowing the coordinator to be easily started and terminated on demand. In practice, coordinators consume minimal resources, as they are not assigned any heavy computational workloads.

\subsubsection*{Worker}

Workers are functions invoked by the coordinator and are responsible for executing graph computations. Each worker retrieves partitioned data and messages from MaaS and performs the required computations. Upon completing a superstep, the worker updates the metadata (e.g., superstep, partition number) in MaaS and waits for the coordinator to initiate the next superstep or finalize the execution. Before termination, each worker writes results back to MaaS.

\subsubsection*{MaaS}

MaaS can be implemented using any storage system that provides efficient and atomic metadata operations. In our Knative deployment, we utilize Dragonfly\footnote{\url{https://www.dragonflydb.io/}} to store control metadata, such as the current superstep, while partitioned graph data is stored in MinIO.\footnote{\url{https://www.min.io/}} Messages may be stored either in the KV store or in object storage, depending on graph size and latency requirements. On AWS, the same design is preserved by replacing MinIO with Amazon S3, which effectively offers horizontally scalable bandwidth that does not become a system-wide bottleneck as the number of workers increases.\footnote{Each Lambda function still faces per-function network limits, but the backend object store itself does not impose centralized bandwidth constraints.} Although Redis-protocol systems naturally support the low-latency atomic primitives used in GraphFlash (e.g., tracking the number of active workers), other storage backends, including transactional SQL databases, can also be adopted as long as they provide the required atomicity and consistency semantics. GraphFlash exposes an abstract MaaS client interface, enabling users to integrate different storage systems depending on their deployment environment and performance requirements.

% MaaS can be implemented using any storage system that provides efficient and atomic metadata operations. In our Knative deployment, we use Dragonfly\footnote{\url{https://www.dragonflydb.io/}} to store control metadata, such as current superstep and small messages, while partitioned graph data and large messages can be placed in MinIO\footnote{\url{https://www.min.io/}}. On AWS, the same design is preserved by replacing MinIO with Amazon S3. Although Redis-protocol systems naturally support low-latency atomic primitives used in GraphFlash (e.g., tracking the number of active workers), other storage backends—including transactional SQL databases—can also be adopted as long as they satisfy the required atomicity and consistency semantics. GraphFlash exposes an abstract MaaS client interface, enabling users to integrate different storage systems depending on their deployment environment.

\subsection{Execution Flow}

\begin{figure}[ht]
    \centering
    \includegraphics[width=\linewidth, trim=0 21 0 17, clip]{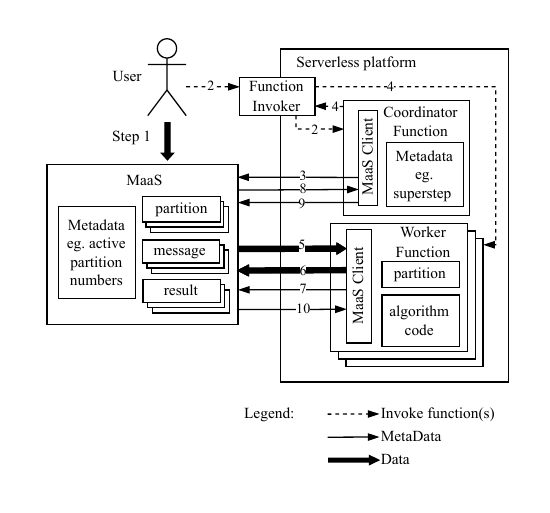}

    \caption{The execution flow of GraphFlash.}
    \label{fig:workflow}
\end{figure}

GraphFlash uses two control flags stored in MaaS: \textit{keep computing}, set by workers to indicate active vertices remain, and \textit{finish}, set by the coordinator to signal termination.
Fig.~\ref{fig:workflow} illustrates the execution flow of a graph processing task in GraphFlash, following the ten steps outlined below. To orchestrate iterative execution,

\begin{enumerate}[leftmargin=*]
    \item Upon receiving the original graph file, users partition it using the high-performance partitioning tool provided by GraphFlash, generating several binary partition files that are then uploaded to MaaS. Users can specify whether the graph is weighted and/or directed to optimize memory usage. Partitions are balanced by total vertex degree.
    
    \item Users send a request to the function invoker to initiate execution, providing necessary information such as the algorithm type as a string (e.g., `BFS`) and the number of partitions. The function invoker may take different forms depending on the deployment platform (e.g., URLs). A coordinator function is then created to manage the entire task.
    
    \item The coordinator writes initial metadata into MaaS, such as the partition count from the request payload, and initializes the superstep to 0.
    
    \item The coordinator invokes the worker functions to process.
    
    \item Upon invocation, each worker loads the required data (e.g., messages and binary partitions) from MaaS and performs computation—PEval in the first superstep and IncVal in subsequent supersteps.
    
    \item After computation, workers write back necessary data to MaaS, including outgoing messages and partial results when required. They terminate if the \textit{finish} flag is set.
    
    \item After persisting all messages and updates, each worker atomically decrements the unfinished-partition counter and may set \textit{keep computing} if active vertices remain.
    
    \item The coordinator polls MaaS until unfinished partitions reach~0, ensuring all workers have completed and persisted outputs; this forms the barrier.
    
    \item If the \textit{keep computing} flag is unset, the coordinator concludes the computation and sets the \textit{finish} flag in MaaS; otherwise, it starts the next superstep and increments the superstep counter to notify the workers.
    
    \item Workers poll MaaS for updates. When the next superstep is available or \textit{finish} is set, they check the metadata to determine whether to proceed with computation or terminate. The process then returns to Step 5.
\end{enumerate}

Steps 5 and 6 involve different data access and write-back patterns depending on the execution mode. GraphFlash provides a parameter called \textit{max\_worker}, which specifies the maximum number of workers allowed to run concurrently. GraphFlash operates in pinned mode when \textit{max\_worker} is greater than or equal to the number of partitions; otherwise, it switches to rotating mode. In other words, pinned mode is used when sufficient resources are available, while rotating mode is designed for resource-constrained environments. Notably, since GraphFlash targets a serverless environment where stateless containers communicate via external storage, there is no strict distinction between push and pull modes. Instead, message generation (push-style) and retrieval (pull-style) coexist in the framework: each container writes intermediate results to MaaS and fetches required data from it in the next step. The details of these two modes are described below.

\subsubsection*{Rotating Mode}

\begin{figure}[ht]
    \centering
    \includegraphics[width=\linewidth, trim=0 5 0 12, clip]{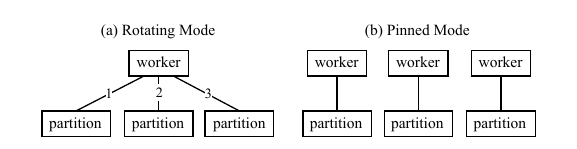}
    \caption{GraphFlash's partition-to-worker assignment in rotating and pinned modes. (a) In rotating mode, a single worker processes multiple partitions in sequence (the numbers indicate access order). (b) In pinned mode, each worker is assigned a fixed partition.}
    \label{fig:flow1}
    
\end{figure}

Rotating mode is the most intuitive approach. Its main characteristic is that in Step 5, workers fetch the partition data from MaaS at every superstep. In addition, starting from superstep 1, workers also retrieve messages and partial results generated in the previous superstep.
Rotating mode follows a concept similar to swap space, allowing it to handle larger datasets with fewer resources. As shown in Fig.~\ref{fig:flow1}, a worker processes a subset of partitions at a time, enabling it to handle multiple subgraphs within a single superstep. The coordinator monitors worker progress and assigns unprocessed partitions to available workers as they complete their tasks.

\begin{figure}[ht]
    \centering
    \includegraphics[width=\linewidth, trim=0 16 0 18, clip]{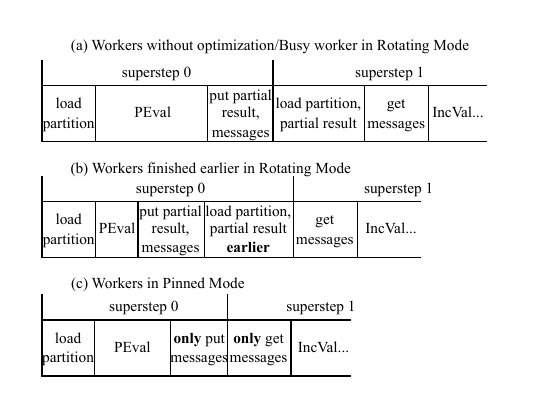}
    \caption{GraphFlash's workers execution timelines in rotating \& pinned modes.}
    \label{fig:flow2}
\end{figure}

In Steps 5 and 6 of Fig.~\ref{fig:workflow}, workers must load partitions, messages, and—if applicable—partial results before performing computation. Rotating mode applies an optimization, illustrated in Fig.~\ref{fig:flow2}. Without this optimization, all workers load their assigned partitions and partial results at the beginning of each superstep. However, near the end of a superstep, if the number of remaining unprocessed partitions is less than \textit{max\_worker}, some workers may finish early and become idle. To improve efficiency, we allow these idle workers to pre-load the partition and partial results for the next superstep in advance.

\subsubsection*{Pinned Mode}

Rotating mode was the initial design. However, during development and testing, we observed that when sufficient functions are available, repeatedly loading partitions becomes unnecessary. This led to the introduction of pinned mode. As shown in Fig.~\ref{fig:flow1}, each partition can be pinned to a dedicated function throughout execution, thereby avoiding both partition loading time and function cold-start latency. In this mode, workers skip partition loading in Step 5. In Step 6, workers only write messages to update external values in other partitions via MaaS, without writing partial results. In the next superstep, they fetch only messages from MaaS, excluding partial results and partitions, as shown in Fig.~\ref{fig:flow2}.

\section{Optimizations}
\label{sec:opt}

To enhance GraphFlash's efficiency, we introduce a series of optimizations specific to our proposed serverless environment. The improvements resulting from the optimizations are reported in Section~\ref{subsec:opti}. 

\subsection{Partition-aware Key Aggregation}

Network communication is a major overhead in serverless environments. In GraphFlash, communication primarily occurs between the function layer and MaaS. Initially, we adopted a per-vertex key design similar to Graphless~\cite{toader2019graphless}, where each vertex is written back as an individual key. As the graph size increases, however, the number of fine-grained operations grows rapidly. In Knative deployments, a high volume of I/O requests can lead to queueing and increased access latency. Even in cloud environments where object stores scale horizontally and do not slow down under higher aggregate load, excessive I/O still accumulates latency on the worker side. Therefore, reducing the number of messages, especially by aggregating fine-grained vertex-level updates, is essential for improving performance.

\begin{figure}[ht]
    \centering
    \includegraphics[width=\linewidth, trim=0 19 0 19, clip]{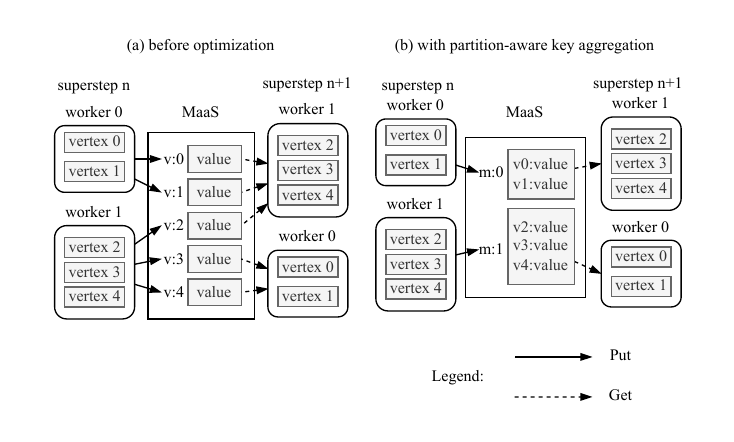}
    \caption{GraphFlash’s partition-aware key aggregation. In this example, the graph has five vertices: worker 0 maintains vertices 0 and 1, while worker 1 maintains the remaining vertices. (a) Before optimization: at the boundary between supersteps $n$ and $n+1$, both workers access all keys in MaaS. (b) With partition-aware key aggregation: both workers access only two keys in MaaS.}
    \label{fig:pa}
    
\end{figure}

To address this, we introduce \textit{partition-aware key aggregation}, as illustrated in Fig.~\ref{fig:pa}. Let $v$ be the total number of vertices and $p$ the number of partitions. Each worker $w_i$ is responsible for partition $P_i$, which contains $v_i$ vertices. In the worst case—when all vertices remain active and depend on remote values in every superstep—a worker may issue up to $v - v_i$ key reads for vertices in other partitions and up to $v_i$ key writes for local updates, resulting in up to $v$ key accesses per superstep (i.e., $\mathcal{O}(v)$).

With partition-aware key aggregation, we reduce the number of keys accessed per worker from $\mathcal{O}(v)$ to $\mathcal{O}(p)$. During graph partitioning, each vertex is annotated with the set of adjacent partitions. At the end of each superstep, worker $w_i$ aggregates all updates destined for partition $P_j$ into a single grouped payload $m_{ij}$, resulting in at most $p-1$ outgoing keys $\{m_{ij}\}_{j \ne i}$. In the next superstep, worker $w_i$ retrieves at most $p-1$ incoming payloads $\{m_{ji}\}_{j \ne i}$. Thus, each worker accesses at most $2(p-1)$ keys per superstep, replacing many fine-grained key-value operations with a small number of partition-level transfers.

\subsection{Intra-function Partition Co-location}

GraphFlash supports multiple partitions per function instance, in contrast to FaaSGraph, where each function is restricted to a single partition. While FaaSGraph adopts shared memory across containers to reduce communication overhead and enable direct data access, this design inherently relies on co-located containers with addressable memory, making it incompatible with serverless environments. Although it exposes a serverless-like API, its architecture violates key serverless principles, such as isolation and statelessness. By contrast, both Graphless and GraphFlash adhere to true serverless constraints.

Graphless follows a one-thread-per-container model and stores all vertex state 
externally. To better utilize CPU resources, GraphFlash assigns one thread per 
partition within a multithreaded function, allowing a single function instance 
to process multiple partitions concurrently. This co-location reduces memory 
consumption by sharing vertex data across partitions and avoiding redundant 
storage of boundary vertices. After remapping, each partition stores its 
vertices in a contiguous memory array for efficient sequential access, while 
co-located partitions can be placed in a unified array to reduce fragmentation. It also allows intra-partition updates to be performed directly in memory without interacting with MaaS.
Intra-partition updates remain local during a superstep and become externally 
visible only after the Bulk Synchronous Parallel (BSP) barrier, fully preserving BSP semantics.

\begin{figure}[ht]
    \centering
    \includegraphics[width=0.9\linewidth, trim=0 16 0 12, clip]{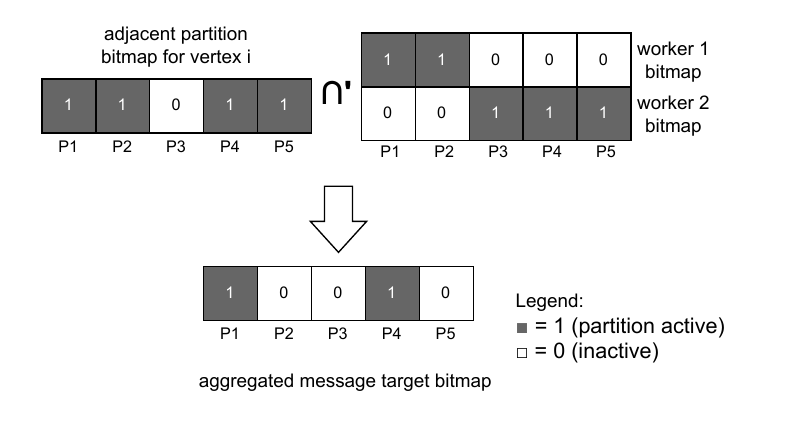}
    \caption{Bitmap-based aggregation to minimize redundant messaging. The symbol $\cap'$ denotes a bitwise masked intersection operation between the vertex’s adjacent partition bitmap and each worker’s partition assignment bitmap.}
    \label{fig:intra}
\end{figure}

Additionally, GraphFlash optimizes inter-function communication by aggregating 
messages at the function level. When a vertex has neighbors in multiple 
partitions that are co-located on the same worker, the message is sent once 
rather than once per partition, avoiding redundant network transfers. To support 
this, we use bitmaps to represent the adjacent partitions of each vertex, and a 
bitmap for each worker indicating its assigned partitions. Using a masked 
bitwise intersection (Fig.~\ref{fig:intra}), GraphFlash determines, for each 
worker, whether any of its partitions require the update.

For example, suppose Vertex~$i$ needs to send updates to Partitions~1, 2, 4, and~5. 
Since Partitions~1 and~2 are co-located on Worker~1, and Partitions~4 and~5 on 
Worker~2, the masked intersection yields one target per worker—Partition~1 for 
Worker~1 and Partition~4 for Worker~2. Consequently, Vertex~$i$ sends only two 
messages instead of four. Because co-located partitions share the same function 
memory region, updating the selected partition implicitly updates its 
co-located peers (e.g., Partitions~1 \&~2, and Partitions~4 \&~5) without extra messaging.

An obvious question is why GraphFlash does not enforce a single partition per function. The reason is flexibility. In serverless systems, resources are 
allocated in fine-grained units (e.g., one vCPU per function), and the optimal 
number of functions depends on workload size, algorithm characteristics, and 
cost-performance tradeoffs. If each function were tied to a single partition, 
changing the function count would require repartitioning the entire graph, an 
expensive and inflexible operation. By allowing a function to process multiple 
partitions, GraphFlash decouples function-level parallelism from the 
partitioning scheme: users can scale the number of functions up or down 
dynamically without modifying the partition layout. This also enables 
memory-heavy algorithms to assign fewer partitions per function, while 
lightweight algorithms can pack more partitions to improve utilization. A 
monolithic one-partition-per-function design cannot provide this elasticity.

\subsection{Superstep-aware Activation}

Active vertex optimization is widely used in distributed graph processing, 
where a vertex is computed only if it has not converged or if any of its 
neighbors have changed. In a serverless environment, however, determining 
vertex activity across partitions is costly: retrieving individual vertex 
values is expensive, and workers rely on pre-aggregated messages rather than 
pulling data on demand. As a result, activation would require explicit 
activation signals from the workers.

In the early supersteps of most algorithms, a large fraction of the vertices 
remain active, making fine-grained activation checks unnecessary and sometimes 
counterproductive due to the additional messaging overhead. GraphFlash therefore allows 
users to enable activation starting from a chosen superstep, avoiding redundant 
checks during the initial phase. 
The appropriate threshold depends on graph density and algorithm behavior. 
% TODO: The threshold can be determined empirically or via lightweight profiling.
For example, dense graphs or slow-converging algorithms (e.g., CDLP) benefit from delayed activation, while sparse graphs or faster-converging ones benefit from enabling it earlier.

\subsection{Other Optimizations}

While JSON is a popular serialization format and is used by Graphless \cite{toader2019graphless}, its human-readable text introduces unnecessary serialization and deserialization latency for GraphFlash. In the context of graph processing, the message content typically consists only of vertex IDs and values, so we directly serialize this information into a binary format. Furthermore, Zstandard (zstd),\footnote{https://www.rfc-editor.org/rfc/rfc8878.pdf} which is a fast lossless compression algorithm,  is used to reduce the size of transferred data and network latency. Since worker functions load serialized partitions from MaaS, large graph sizes can cause memory usage to increase significantly when vertex values are naively serialized and compressed. This is critical due to the memory constraints of serverless functions. To address this, we apply prefix compression and varint encoding. Specifically, when serializing outgoing edges, we sort the destination vertex IDs, record the first ID explicitly, and encode subsequent IDs as differences using varints. This reduces the serialized data size compared to storing each vertex ID as a 4-byte integer. During Knative testing, we also identified bottlenecks in the Container Network Interface~(CNI), which slowed GraphFlash at the beginning and end of supersteps due to inefficiencies in network packet processing. To address this, we employ message batching to reduce the number of network packets between functions and MaaS, and adopt Cilium,\footnote{https://cilium.io/} a high-performance CNI solution, to further improve networking throughput and scalability.

\section{Evaluation}
\label{sec:evaluation}

\subsection{Implementation and Experimental Setup}

GraphFlash is implemented in C++ and released as open source, with two versions: one for Knative and one for AWS Lambda.\footnote{
\url{https://github.com/disnetlab/GraphFlash/tree/v1.0-knative} (Knative), 
\url{https://github.com/disnetlab/GraphFlash/tree/v1.0-lambda} (AWS Lambda)
} In the Knative version, functions are implemented as lightweight HTTP servers with a single request handler, and autoscaling is set to 1 to ensure one request per instance. For the AWS Lambda version, we build the image using the AWS custom runtime.

% In the Knative version, functions are implemented as lightweight HTTP servers, each with a single handler to process requests. The autoscaling parameter is set to 1, ensuring that each HTTP request is handled by an individual function instance, enabling fine-grained control over resource management. For the AWS Lambda version, we build the image based on the custom runtime provided by AWS.

Most experiments are conducted on a cluster comprising four nodes, each equipped with a 32-core AMD EPYC 9474F CPU, 128GB of DRAM, and 25Gbps inter-node bandwidth. The nodes share a high-performance NFS with read and write speeds of approximately 215MB/s and 205MB/s, respectively. On the Knative cluster, MaaS consists of three 4-thread Dragonfly servers for storing metadata, messages, and results with low latency, and one MinIO server for partitioned data to achieve high throughput. The full hardware and software configuration is listed in Table~\ref{tab:impl}. We also evaluated GraphFlash on AWS Lambda to compare with the existing serverless-compatible solutions.

\begin{table}[ht]
  \caption{Hardware and Software Configurations}
  \label{tab:impl}
  \scriptsize
  \begin{tabular}{p{1cm}p{6.5cm}}
  \toprule
  & Configuration\\
  \midrule
  Hardware & CPU: AMD EPYC 9474F @3.6GHz; Cores: 32, DRAM: 128GB; Inter-node Bandwidth: 25Gbps; Storage: Shared NFS (read: $\sim$215MB/s; write: $\sim$205MB/s) \\
  Software & Ubuntu 24.04.5 LTS, g++ 13.3.0; Kubernetes 1.30.8, Knative Serving 1.16.0; Docker 28.0.4, Dragonfly 1.62.0; MinIO RELEASE.2024-12-18T13-15-44Z\\
  \bottomrule
  \end{tabular}
\end{table}

\subsubsection{Datasets and Graph Algorithms}

We evaluate GraphFlash using several representative datasets from the commonly used 
LDBC Graphalytics benchmark, following dataset scale settings in prior work 
\cite{toader2019graphless, 10.1145/3620665.3640361, fan2021graphscope}. 
The datasets include both real-life and synthetic graphs. 
For real-life graphs, we use dota-league (\textbf{DL}) with 61.1K vertices and 50.9M edges, 
and com-friendster (\textbf{CF}) with 65.6M vertices and 1.81B edges. 
For synthetic graphs, we use graph500-23 (\textbf{G3}) with 4.61M vertices and 129.3M edges, 
graph500-25 (\textbf{G5}) with 17.1M vertices and 523.6M edges, 
graph500-26 (\textbf{G6}) with 32.8M vertices and 1.15B edges, 
graph500-27 (\textbf{G7}) with 63.1M vertices and 2.11B edges, 
datagen-9\_2-zf (\textbf{ZF}) with 434.9M vertices and 1.04B edges, 
and graph500-28 (\textbf{G8}) with 121.2M vertices and 4.23B edges. 
These datasets span diverse graph regimes, including  vertex-heavy graphs (e.g., ZF), and  heavy graphs with high communication demands (e.g., G8), covering a wide range of sizes and connectivity patterns. 
Such diversity allows us to evaluate GraphFlash comprehensively across 
different workload characteristics and structurally varied graphs.

We use the same hardware and dataset configurations consistently throughout the performance optimization and evaluation experiments presented in subsequent sections. Each dataset is evaluated using four widely adopted graph algorithms: (a) \textbf{BFS} (Breadth-First Search), which traverses the graph layer by layer from a given source vertex, visiting all reachable vertices; (b) \textbf{PageRank}, which iteratively computes an importance score for each vertex based on the structure of incoming links and is widely used in ranking tasks; (c) \textbf{CDLP} (Community Detection using Label Propagation), which identifies communities by iteratively propagating labels among neighboring vertices in parallel; and (d) \textbf{WCC} (Weakly Connected Components), which determines the connected component each vertex belongs to in a directed graph, treating edges as undirected.

\subsubsection{Baselines and Benchmarks}

In addition to evaluating GraphFlash in its two execution modes, we compare it against two serverless frameworks and two conventional graph processing systems. A brief description of each baseline is as follows:

%\begin{itemize} 

%\item 
\noindent\textbf{--Graphless} \cite{toader2019graphless}: A serverless graph processing framework that maps vertices to stateless functions, using a distributed key-value store for state and external storage for scalability.

\noindent\textbf{--FaaSGraph}\cite{10.1145/3620665.3640361}: A serverless-inspired framework that utilises shared memory and proxy mechanisms to reduce communication overhead, not deployable on pure serverless infrastructure, as it depends on in-memory processing and co-located resources for efficiency. % Specifically, FaaSGraph relies on shared memory across co-located containers and host-level proxies, which renders it incompatible with true serverless deployment and prevents a fair assessment of cost.
To enable fair comparison, we executed FaaSGraph in Docker containers on dedicated machines, following the deployment procedure provided.

\noindent\textbf{--GraphScope}\cite{fan2021graphscope}: A distributed graph processing system that integrates multiple engines to support both iterative graph algorithms and machine learning tasks. 

\noindent\textbf{--Giraph}\cite{averyGiraph}: A classical graph processing framework that implements the BSP model through iterative supersteps, deployed in Hadoop-based environments. 

%\end{itemize}

%We also attempted to evaluate earlier systems such as PowerGraph~\cite{10.5555/2387880.2387883},
%PowerLyra~\cite{10.1145/3298989}, and Gemini~\cite{10.5555/3026877.3026901}. 
%PowerGraph and PowerLyra rely on outdated software stacks that are no longer 
%maintained, while Gemini requires NUMA hardware unavailable on our testbed. 
%These systems are therefore omitted from our evaluation.

% We also made efforts to evaluate several other earlier systems, including PowerGraph~\cite{10.5555/2387880.2387883}, PowerLyra~\cite{10.1145/3298989}, and Gemini~\cite{10.5555/3026877.3026901}, However, these systems could not be executed effectively in our environment due to factors such as lack of ongoing maintenance, reliance on outdated software dependencies, specific hardware constraints (e.g., the NUMA-awareness requirement in Gemini). %These systems have made significant contributions to the field, but are no longer suitable for comparative evaluation under contemporary deployment environments. 
%In contrast, the selected baselines are either actively maintained or recently updated:
%GraphScope remains under active development,
%FaaSGraph was introduced in 2024 and continues to be maintained,
%Giraph was last updated in 2021, and
%Graphless, although introduced in 2019, represents the earliest serverless design in this space and is included for historical comparison.

We measure the execution time of graph algorithms on the datasets across all frameworks in our cluster. For consistency, Graphless is ported to our Knative setup. GraphFlash is also deployed on AWS Lambda to enable direct comparison with the official Graphless deployment. For cost evaluation, we use core$\cdot$seconds and GB$\cdot$seconds on Knative, and actual monetary cost on AWS Lambda based on its billing model.\footnote{AWS Lambda x86-based functions (up to the first 6 billion GB-seconds per month) are priced at \$0.0000166667/GB-second and \$0.20/million requests.}

\subsection{Experimental Results and Analysis}
\subsubsection{Evaluation of Execution Times}

\begin{figure}[ht]
    \centering
    \includegraphics[width=\linewidth]{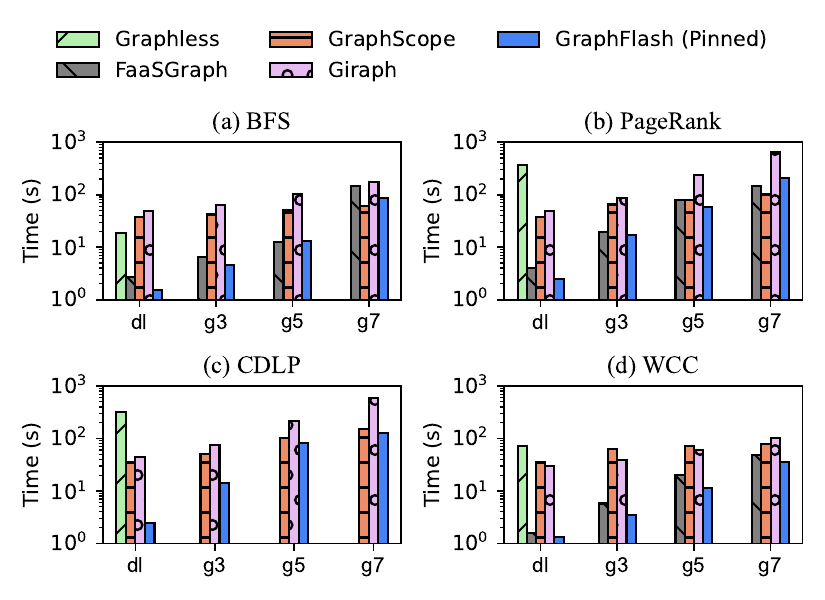}
    \caption{Performance of frameworks across algorithms and datasets (total execution time). Graphless runs only on the DL due to its limitation, and CDLP is not tested on FaaSGraph since the corresponding implementation is not available. Reported times exclude one-time costs such as compilation and deployment.}
    \label{fig:e-performance}
\end{figure}

We evaluate the execution times of GraphFlash (pinned mode) in comparison with both serverless and conventional distributed graph processing frameworks, including Graphless, FaaSGraph, GraphScope, and Giraph. The evaluation was conducted on the cluster configuration described in Table~\ref{tab:impl}, using DL, G3, G5, and G7. Fig.~\ref{fig:e-performance} presents the execution time comparisons.

For smaller graphs such as DL and G3, GraphFlash outperforms all other frameworks across all supported algorithms. GraphScope, which is not a serverless framework, despite having efficient execution engines, suffers from long pre-compilation times that introduce significant overhead, resulting in slower performance on smaller datasets. As the graph size increases, the relative impact of this compilation overhead diminishes, and GraphScope's performance becomes comparable to that of GraphFlash, particularly on large datasets such as G7, due to increased message volume in GraphFlash and a higher partition count, which results in more outer vertices and communication overhead.

For larger graphs such as G5 and G7, GraphFlash achieves execution times comparable to or better than GraphScope and FaaSGraph, depending on the specific algorithm, while generally outperforming Giraph. Due to its architectural limitations, Graphless can only run on the smallest dataset, DL. In addition, since FaaSGraph does not support the CDLP algorithm, the results for these cases are omitted. Among frameworks designed for serverless environments, Graphless is slower than the classic distributed framework, i.e., Giraph, whereas GraphFlash overcomes this limitation and achieves competitive or superior performance. Compared to Graphless, GraphFlash is at least 12$\times$ faster and achieves a speedup of up to 127$\times$ on certain algorithms, e.g.,  CDLP.

In summary, GraphFlash presents a viable and efficient approach to serverless graph processing, especially for small to medium datasets with dynamic and bursty workloads. On large-scale workloads, it achieves comparable performance while offering the flexibility and scalability of serverless deployment. %By supporting larger datasets and reducing preprocessing overhead, GraphFlash extends the practical applicability of serverless paradigms to more demanding graph workloads.

\subsubsection{Evaluation of Cost Efficiency}

\begin{figure*}[ht]
    \centering
    \includegraphics[width=\linewidth]{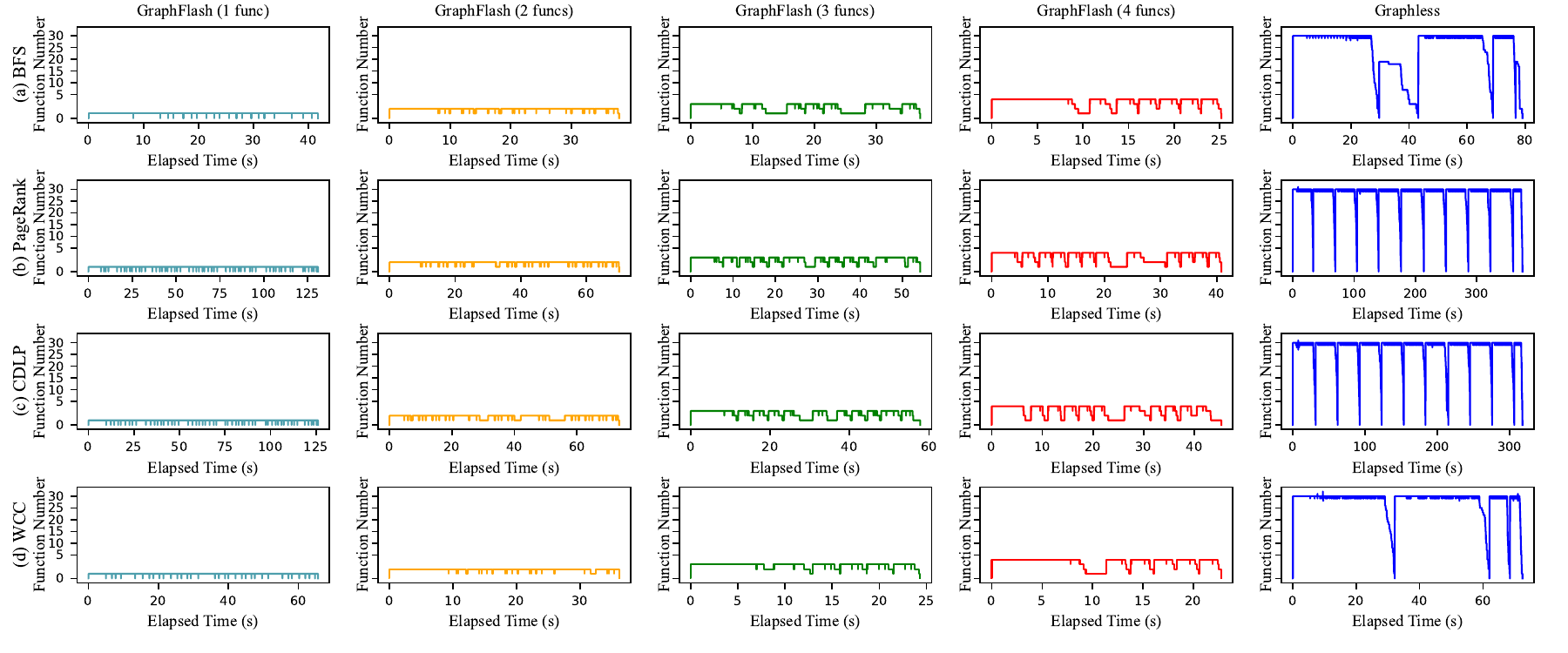}
    \vspace{-0.8cm}
    \caption{Number of running functions over time: comparison between GraphFlash (rotating mode) and Graphless.}
    \label{fig:cpumem}
\end{figure*}

We compare GraphFlash and Graphless, the only two frameworks that can run on serverless platforms, and for which the cost is fair and straightforward to measure. Note that FaaSGraph is not included, as it relies on shared memory across co-located containers with host-level proxies, which makes it incompatible with true serverless deployment and prevents a fair accounting of memory usage and cost. Each function is allocated 1 core and 2GB of memory, an appropriate configuration for both frameworks, and we run them on the DL dataset. The number of concurrently running functions during execution is recorded and shown in Fig.~\ref{fig:cpumem}. In the case of GraphFlash, we partition the input graph into five parts and run in rotating mode with \textit{max\_worker} values ranging from 1 to 4. As expected, increasing \textit{max\_worker} reduces the overall execution time. All configurations of GraphFlash complete faster than Graphless, while requiring significantly fewer concurrent functions. 
% The results show that GraphFlash can handle a graph containing 50 million edges with only one function across five partitions, which is not achievable by FaaSGraph. 
When running a single function, GraphFlash delivers substantial cost savings, reducing resource consumption by 96.8\% for WCC and up to 98.8\% for PageRank compared with Graphless.

\subsubsection{Effect of Partition Count on Performance}

\begin{figure}[ht]
    \centering
    \includegraphics[width=\linewidth, trim=0 12 0 15, clip]{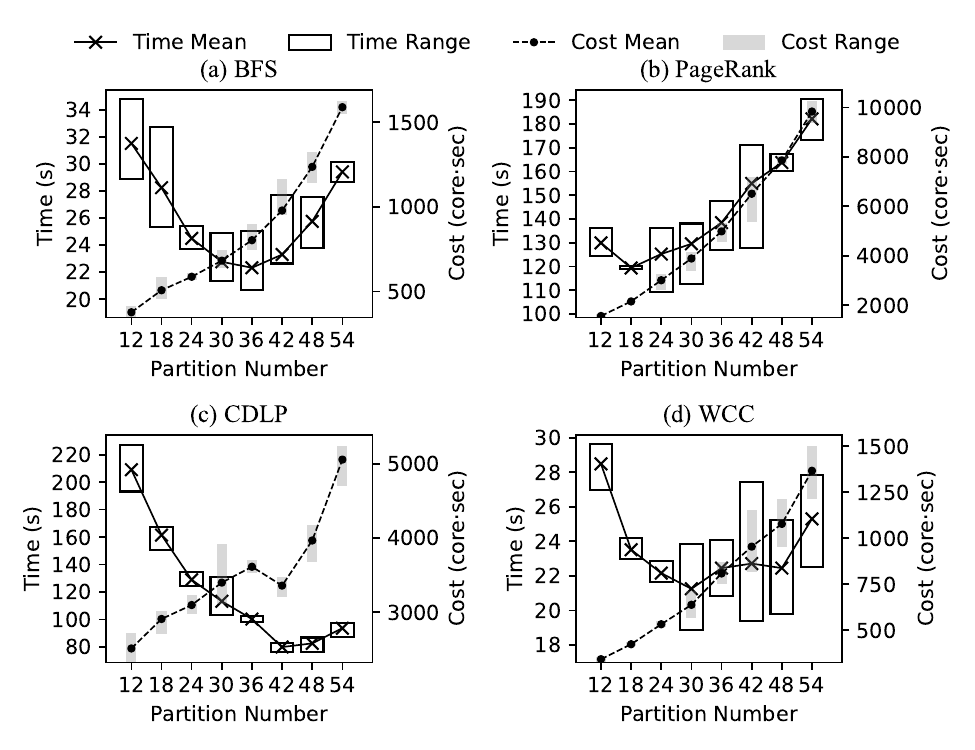} 
    \caption{Execution times and costs of pinned mode across different partition numbers on input graph G6. The shaded regions represent the range (min to max) over multiple runs, and the lines with markers indicate the average values.}
    \label{fig:partition_time}
    \vspace{-0.8cm}
\end{figure}

In this subsection, we explore the impact of partition count on execution time and cost.
As mentioned earlier, GraphFlash can achieve higher parallelism with more partitions; however, this also leads to larger edge cuts and increased message exchange, which may degrade performance. We evaluate this trade-off using the G6 dataset by executing the four algorithms with partition numbers ranging from 12 to 54. The results are presented in Fig.~\ref{fig:partition_time}.

We observe that for different algorithms, the optimal partition number—that is, the one yielding the lowest execution time—varies. For example, PageRank reaches its minimum execution time at 18 partitions, however, performance degrades quickly beyond that due to increased time spent generating and exchanging messages, as edge cuts grow. In contrast, CDLP, which involves more intensive computation, achieves optimal performance at 42 partitions, later than the other algorithms.

An interesting observation is that the cost (core $\cdot$ sec) increases consistently as the number of partitions grows. This is attributed to the rising edge cut ratio, which results in more time spent on inter-function message exchange. Although network communication is not the system bottleneck, workers must generate a larger number of messages at the end of each superstep and process more incoming messages at the beginning of the next, due to the increased number of outer vertices introduced by finer partitioning.

\subsubsection{Evaluation on AWS Lambda}
\label{subsec:lambda}

We evaluate GraphFlash on AWS Lambda in two parts. First, we use the DL dataset 
to directly compare against Graphless, following the same setup reported in its 
paper. Since DL is relatively small, a single 768MB Lambda function is 
sufficient to process the entire graph. This allows GraphFlash to exploit 
intra-function partition co-location while keeping invocation overhead low. 
To ensure fairness, we treat our 768MB function as equivalent to six 128MB 
functions, matching Graphless in total memory allocation and cost.\footnote{
AWS Lambda allocates approximately one vCPU per 1769MB of memory.}
Execution time and function usage for Graphless are taken directly from the 
published results, as per-function memory configurations were not specified.

\begin{figure}[ht]
    \centering
    \includegraphics[width=\linewidth]{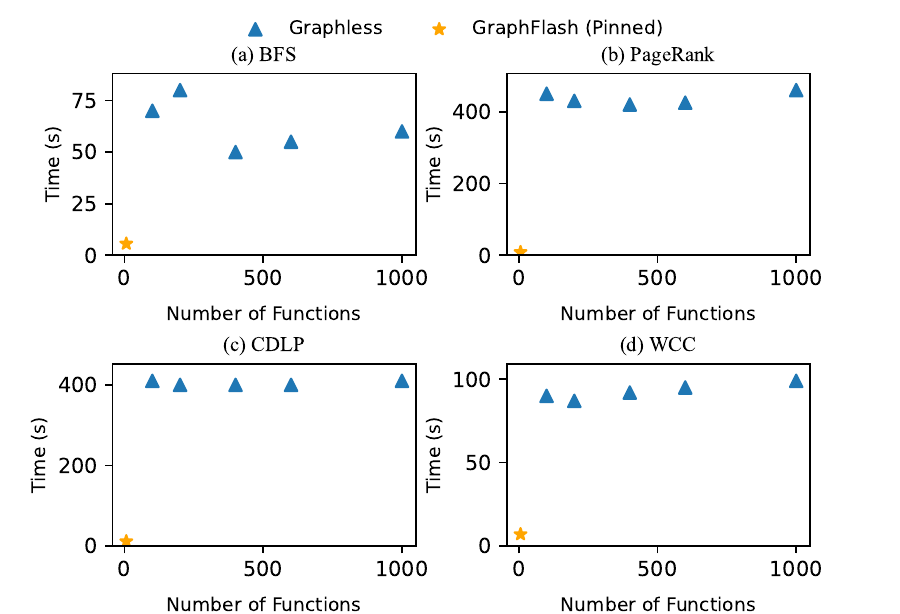}
    \caption{Execution time (s) and function number comparison with Graphless on AWS Lambda on graph DL.}
    \label{fig:lambdatime}
\end{figure}

\begin{figure}[ht]
    \centering
    \includegraphics[width=\linewidth, trim=0 12 0 10, clip]{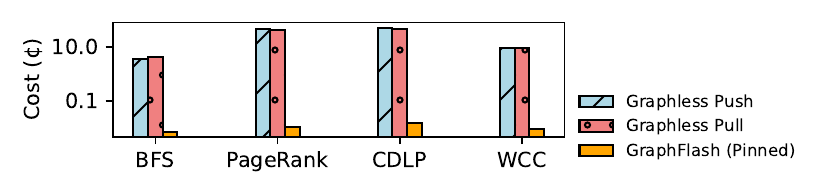}
    \caption{Cost (\textcent) comparison with Graphless on AWS Lambda on graph DL.}
    \label{fig:lambdacost}
    \vspace{-0.3cm}
\end{figure}

As shown in Fig.~\ref{fig:lambdatime} and Fig.~\ref{fig:lambdacost}, 
GraphFlash achieves up to 48$\times$ speedup and 99.97\% cost reduction over 
Graphless on DL, despite each function receiving only 0.4~vCPU. Even in this 
resource-constrained setting, GraphFlash provides 9$\times$ speedup for BFS and 
over 48$\times$ for PageRank.

Beyond DL, Graphless cannot be executed on larger datasets, so we evaluate 
GraphFlash independently on CF, ZF, and G8 to assess scalability. Data as shown in 
Table~\ref{tab:lambda_scalability}, GraphFlash maintains stable performance 
across increasing graph sizes and parallelism levels, demonstrating that the 
system design scales robustly on AWS Lambda. This result reflects the combined 
effect of the optimizations introduced in Section~\ref{sec:opt}, which together 
enable GraphFlash to operate efficiently even on very large datasets.

Together, these results demonstrate that GraphFlash not only outperforms 
Graphless on DL, but also scales reliably to much larger datasets on AWS Lambda.

\begin{table}[t]
\centering
\caption{Execution time (s) of GraphFlash on AWS Lambda across datasets of 
different sizes, demonstrating scalability.}
\label{tab:lambda_scalability}
\small
\begin{tabular}{lcccc}
\toprule
Dataset & BFS & PageRank & CDLP & WCC \\
\midrule
CF (64 partitions)
& 102.95 & 154.14 & 281.94 & 126.02 \\
ZF (256 partitions)
& 226.46 & 328.18 & 321.48 & 241.83 \\
G8 (256 partitions)
& 167.07 & 311.61 & 321.48 & 199.87 \\
\bottomrule
\end{tabular}
\end{table}

\subsubsection{Ablation Study}
\label{subsec:opti}

We ablate each optimization from Section~\ref{sec:opt}, using the ablated variant as baseline and the fully optimized GraphFlash as reference.
Speedup ($\times$) is defined as the ratio of baseline execution time to optimized execution time. All ablation experiments in this section are conducted on the Knative deployment.

%\subsubsection*{Partition-Aware Key Aggregation}

We first ablate \textit{partition-aware key aggregation} by disabling it and measuring the resulting execution times on DL, G3, and G5. To avoid interference from intra-function partition co-location, the baseline here is configured with one function per partition, so that no partitions are colocated. 
As shown in Table~\ref{tab:keyaggregation}, the largest gains are observed in larger graphs and for CDLP and WCC.

\begin{table}[t]
\caption{Execution time improvement with partition-aware key aggregation. \textit{Speedup} is how many times faster the optimized version is than the baseline.}
  \label{tab:keyaggregation}
\centering
\small
\begin{tabular}{clccc}
\toprule
Dataset & Algorithm & Baseline(s) & Optimized(s) & Speedup \\
\midrule
\multirow{4}{*}{DL} 
&BFS & 3.29 & 2.26 & 1.46 \\
&PageRank & 9.01 & 3.76 & 2.40\\
&CDLP & 14.2 & 7.56 & 1.88 \\
&WCC & 4.63 & 3.71 & 1.25 \\
\midrule
\multirow{4}{*}{G3} 
&BFS & 17.2 & 6.68 & 2.57 \\
&PageRank & 30.5 & 23.4 & 1.30 \\
&CDLP & 128 & 36.6 & 3.50 \\
&WCC & 20.7 & 5.43 & 3.81 \\
\midrule
\multirow{4}{*}{G5} 
&BFS & 26.9 & 13.1 & 2.05 \\
&PageRank & 139 & 80.9 & 1.72 \\
&CDLP & 484 & 119 & 4.06 \\
&WCC & 103 & 19.0 & 5.42 \\
\bottomrule
\end{tabular}
\end{table}

%\subsubsection*{Intra-function Partition Co-location}

Next, we examine the effect of \textit{intra-function partition co-location} by varying the number of vCPUs per function (1 vs. 6) on Knative while keeping the total vCPUs fixed.\footnote{According to the AWS Lambda specification, vCPUs scale with memory allocation—approximately one vCPU per 1,769MB. For example, a function with 10,240MB gets 6 vCPUs. Thus, the cost of one function with 6 vCPUs is equivalent to that of six functions with 1 vCPU.} GraphFlash automatically derives the number of functions based on the partition count and the vCPU allocation per function. As shown in Table~\ref{tab:thread_vs_time} and Table~\ref{tab:thread_vs_mem}, this optimization reduces both execution time and memory usage. The gains are substantial for all algorithms on G3, and remain noticeable for BFS, PageRank, and WCC on G6. Memory usage is significantly reduced (by over 50\%) due to shared storage of adjacent vertices and reusable buffers. This is especially valuable for large graphs, where duplicated data can otherwise increase memory consumption.

Finally, we ablate \textit{superstep-aware activation}. Table~\ref{tab:activation_opt} reports the results, showing modest but consistent improvements across all algorithms, with WCC achieving up to 25\% reduction in execution time.

\begin{table}[ht]
\caption{Execution time (s) with different thread numbers in each function ($t$), with a fixed total resource budget. G3 and G6 are partitioned into 6 and 24 partitions, respectively.}
  \label{tab:thread_vs_time}
\centering
\small
\begin{tabular}{clccc}
\toprule
Dataset & Algorithm & Time($t=1$) & Time($t=6$) & Speedup \\
\midrule
\multirow{4}{*}{G3}
&BFS & 5.09 & 3.86 & 1.32 \\
&PageRank & 19.2 & 17.9 & 1.07 \\
&CDLP & 26.9 & 20.1 & 1.34 \\
&WCC & 4.86 & 3.71 & 1.31 \\
\midrule
\multirow{4}{*}{G6}
&BFS & 30.0 & 24.5 & 1.22 \\
&PageRank & 89.1 & 83.5 & 1.08 \\
&CDLP & 99.1 & 98.9 & 1.00 \\
&WCC & 19.7 & 17.6 & 1.12 \\
\bottomrule
\end{tabular}
\vspace{-0.3cm}
\end{table}

\begin{table}[ht]
\caption{Memory usage with different thread numbers in each function ($t$).}
  \label{tab:thread_vs_mem}
\centering
\small
\begin{tabular}{cccc}
\toprule
Dataset & Memory($t=1$) & Memory($t=6$) & Saving(\%) \\
\midrule
G3 & 543MB & 254MB & 53.2 \\
G6 & 52.3GB & 23.2GB & 55.6 \\
\bottomrule
\end{tabular}
\end{table}

%\subsubsection*{Superstep-Aware Activation}

\begin{table}[ht]
\caption{Execution time improvement with superstep-aware activation. G3 and G6 are partitioned into 6 and 24 partitions, respectively.}
  \label{tab:activation_opt}
\centering
\small
\begin{tabular}{clccc}
\toprule
Dataset & Algorithm & Baseline(s) & Optimized(s) & Speedup \\
\midrule
\multirow{3}{*}{G3}
&BFS & 4.04 & 3.86 & 1.05 \\
&CDLP & 22.4 & 20.1 & 1.11 \\
&WCC & 4.63 & 3.71 & 1.25 \\
\midrule
\multirow{3}{*}{G6}
&BFS & 26.7 & 24.5 & 1.09 \\
&CDLP & 105 & 98.9 & 1.06 \\
&WCC & 21.8 & 17.6 & 1.23 \\
\bottomrule
\end{tabular}
\vspace{-0.3cm}
\end{table}

\section{Related Work}
\label{sec:related}

%In this section, we present an overview of related work.

\textbf{Distributed Graph Processing.} GraphFlash is part of the family of distributed graph processing systems, a field with numerous existing solutions %\cite{averyGiraph,10.1145/2484425.2484427,10.1145/3298989,10.5555/3026877.3026901,10.5555/2387880.2387883,10.1145/2487575.2487581,10.1145/2807591.2807620,10.1145/2524211.2524218,10.1145/2465351.2465369,10.5555/2387880.2387884,toader2019graphless,fan2021graphscope,10.1145/3620665.3640361,10.1145/1807167.1807184,10.1145/2517349.2522739,10.1145/2815400.2815408,10.1145/2517349.2522740,10.1145/2484838.2484843,5708522,10.1145/2517327.2442530,10.14778/2809974.2809983,wu2015gram,7013020,zheng2015flashgraph,10.5555/2813767.2813795}.
Unlike most distributed graph processing systems that require users to manage machines and adjust clusters to handle varying dataset sizes, only a few, such as Graphless \cite{toader2019graphless} and FaaSGraph \cite{10.1145/3620665.3640361}, offer a serverless approach. GraphFlash introduces a more flexible, serverless solution for graph processing. Compared with Giraph \cite{averyGiraph} and GraphScope \cite{fan2021graphscope}, GraphFlash eliminates the need for users to manage machine allocation, cluster scaling, or termination, simplifying operations while delivering high performance on small datasets and competitive performance on medium-sized datasets. As such, GraphFlash offers a cost-effective and operationally lightweight alternative for processing graphs that do not require large-scale cluster resources.

\textbf{Serverless Data Processing.} Serverless computing has demonstrated significant potential in data processing; however, most existing systems are not designed to run graph algorithms or support graph processing tasks. ServerlessLLM \cite{298681} is a serverless distributed system optimized for low-latency inference of Large Language Models (LLMs), leveraging near-GPU storage, fast checkpoint loading, live migration, and model scheduling optimized for startup time. Starling \cite{10.1145/3318464.3380609} is a query execution engine built on serverless cloud function services, tailored to handle bursty or low-volume database analytics workloads with lower costs and interactive latency. Lambada~\cite{10.1145/3318464.3389758} is a serverless distributed data processing framework that examines the economic and performance trade-offs of serverless computing for data analytics across various domains. Cirrus~\cite{10.1145/3357223.3362711} is a serverless machine learning (ML) framework that automates resource management for end-to-end ML workflows, combining the simplicity of serverless interfaces with scalability to optimize iterative ML training.

Graphless~\cite{toader2019graphless} is the pioneering serverless graph processing system. However, its reliance on vertex-centric programming models and the absence of optimizations tailored for serverless environments lead to performance bottlenecks, primarily caused by communication overhead. On the other hand, FaaSGraph~\cite{10.1145/3620665.3640361} introduces a data-centric execution model and incorporates several optimizations to address the challenges of serverless computing, resulting in high performance. However, its components run on Docker and require manual orchestration. Additionally, components designed to accelerate processing, such as proxies, need to be addressable, which conflicts with the stateless nature of functions in serverless computing, making it challenging to fully migrate to serverless platforms. Furthermore, it requires sufficient resources to run functions together and thus does not support processing large datasets with limited resources. Table~\ref{tab:framework_comparison} summarizes key differences among mainstream graph processing frameworks, highlighting the advantages of GraphFlash in serverless compatibility, performance, and efficiency.

\begin{table}[t]
\centering
\caption{Comparison of Graph Processing Frameworks.}
\label{tab:framework_comparison}
\small
\resizebox{\linewidth}{!}{
\begin{tabular}{@{}lccccc@{}}
\toprule
Feature & GraphFlash & FaaSGraph & Graphless & GraphScope & Giraph \\ \midrule
Serverless Compatible & \cmark & \xmark & \cmark & \xmark & \xmark \\
Performance & High & High & Low & High & Medium \\
Cost Efficiency & High & N/A & Low & N/A & N/A \\
\bottomrule

\end{tabular}
}
\vspace{-0.3cm}
\end{table}

\section{Conclusion}
\label{sec:conclusion}
We propose GraphFlash, a practical and fully serverless graph processing framework that combines elasticity with high performance. By identifying key challenges, such as communication overhead, memory duplication, and inefficient early-stage execution, we design optimizations tailored for serverless environments, including partition-aware key aggregation, intra-function partition co-location, and superstep-aware activation.
Our evaluation shows that GraphFlash achieves significant performance improvements over existing serverless-compatible frameworks. On small to medium datasets, it outperforms both serverless and traditional systems. On large-scale workloads, it delivers execution times comparable to high-performance frameworks such as FaaSGraph and GraphScope, while preserving full compatibility with serverless infrastructure.
Importantly, GraphFlash offers substantial cost advantages, reducing resource usage by $>$90\% compared to prior serverless designs. It also features a lightweight deployment model with a minimal container image size, enabling fast scaling and reduced overhead in cloud environments.
In future work, we aim to design serverless platforms that support direct function-to-function communication, eliminating reliance on MaaS for intermediate messaging to further reduce latency and improve performance.

\section*{Acknowledgments}
The authors thank the Melbourne Research Cloud at the University of Melbourne for providing computational resources.

\bibliographystyle{IEEEtran}
\bibliography{references}

% \begin{thebibliography}{00}
% \bibitem{b1} G. Eason, B. Noble, and I. N. Sneddon, ``On certain integrals of Lipschitz-Hankel type involving products of Bessel functions,'' Phil. Trans. Roy. Soc. London, vol. A247, pp. 529--551, April 1955.
% \bibitem{b2} J. Clerk Maxwell, A Treatise on Electricity and Magnetism, 3rd ed., vol. 2. Oxford: Clarendon, 1892, pp.68--73.
% \bibitem{b3} I. S. Jacobs and C. P. Bean, ``Fine particles, thin films and exchange anisotropy,'' in Magnetism, vol. III, G. T. Rado and H. Suhl, Eds. New York: Academic, 1963, pp. 271--350.
% \bibitem{b4} K. Elissa, ``Title of paper if known,'' unpublished.
% \bibitem{b5} R. Nicole, ``Title of paper with only first word capitalized,'' J. Name Stand. Abbrev., in press.
% \bibitem{b6} Y. Yorozu, M. Hirano, K. Oka, and Y. Tagawa, ``Electron spectroscopy studies on magneto-optical media and plastic substrate interface,'' IEEE Transl. J. Magn. Japan, vol. 2, pp. 740--741, August 1987 [Digests 9th Annual Conf. Magnetics Japan, p. 301, 1982].
% \bibitem{b7} M. Young, The Technical Writer's Handbook. Mill Valley, CA: University Science, 1989.
% \end{thebibliography}
\end{document}